\documentclass{aa}
\usepackage{txfonts}
\usepackage{graphicx}
\usepackage{subcaption}
\usepackage[colorlinks=true,allcolors=blue]{hyperref}

\begin{document} 

\titlerunning{}
\authorrunning{Mountrichas George}

\title{The coevolution of supermassive black holes and galaxies in luminous AGN over a wide range of redshift}

\author{George Mountrichas\inst{1}}
          
     \institute {Instituto de Fisica de Cantabria (CSIC-Universidad de Cantabria), Avenida de los Castros, 39005 Santander, Spain
              \email{gmountrichas@gmail.com}
              }

\abstract{It is well known that supermassive black holes (SMBHs) and their host galaxies co-evolve. A manifestation of this co-evolution is the correlation that has been found between the SMBH mass, M$_{BH}$, and the galaxy bulge or stellar mass, M$_*$. The cosmic evolution of this relation, though, is still a matter of debate. In this work, we examine the M$_{BH}-$M$_*$ relation, using 687 X-ray luminous (median $\rm log\,[L_{X,2-10keV}(ergs^{-1})]=44.3$), broad line AGN, at $\rm 0.2<z<4.0$ (median $\rm z\approx 1.4$) that lie in the XMM-{\it{XXL}} field. Their M$_{BH}$ and M$_*$ range from $\rm 7.5<log\,[M_{BH}\,(M_\odot)]<9.5$ and $\rm 10<log\,[M_*(M_\odot)]<12$, respectively. Most of the AGN live in star-forming galaxies and their Eddington ratios range from 0.01 to 1, with a median value of 0.06. Our results show that M$_{BH}$ and M$_*$ are correlated ($\rm r=0.47\pm0.21$, averaged over different redshift intervals). Our analysis also shows that the mean ratio of the M$_{BH}$ and M$_*$ does not evolve with redshift, at least up to $\rm z=2$ and has a value of $\rm log($M$_{BH}/$M$_*)=-2.44$. The majority of the AGN ($75\%$) are in a SMBH mass growth dominant phase. In these systems, the M$_{BH}-$M$_*$ correlation is weaker and their M$_*$ tends to be lower (for the same M$_{BH}$) compared to systems that are in a galaxy mass growth phase. Our findings suggest that the growth of black hole mass occurs first, while the early stellar mass assembly may not be so efficient.}



   
\maketitle

\section{Introduction}

In the last two decades, several studies have shown that there is a co-evolution between the supermassive black holes (SMBHs) and their host galaxies \citep[e.g.,][]{Kormendy2013}. In the local universe, this co-evolution has been demonstrated by tight correlations that have been found between the SMBH mass, M$_{BH}$, and various properties of the host galaxy. For instance, there is a correlation between the M$_{BH}$ and the stellar velocity dispersion, the bulge luminosity and the bulge mass, M$_{bulge}$ \citep[e.g.,][]{Magorrian1998, Ferrarese2000, Gebhardt2000, Tremaine2002, Haring2004}. Among them, the correlation between the M$_{BH}$ and the velocity dispersion of the galaxy bulge ($\sigma$) appears to be the tightest. A possible explanation could be that $\sigma$ is a good predictor of M$_{bulge}$. Another, perhaps more plausible, scenario is that $\sigma$ measures the depth of the potential well in which the SMBH is formed \citep{Ferrarese2000}. 

Although these correlations are well established at low redshift ($\rm z<1$), it is still unclear if and how they evolve at high redshifts. A comparison of the local scaling relations with those at higher redshifts is not straightforward. \cite{Shankar2016} used Monte Carlo simulations and found evidence that local galaxy samples with dynamically measured M$_{BH}$ may suffer from an angular resolution related selection effect that could bias the observed scaling relations between the M$_{BH}$ and galaxy properties. However, this selection effect does not affect local samples of active galactic nuclei \citep[AGN;][]{Shankar2019}. 

The M$_{BH}$ and M$_{bulge}$ or stellar mass, M$_*$ is one of the most extensively studied relations, both from a theoretical as from an observational point of view \citep[e.g.][]{Marconi2003, Gueltekin2009, Sani2011, Reines2015}. Since, at high redshift, it is difficult to separate the bulge from the total stellar mass, many observational works at $\rm z \geq 1$ have studied the M$_{BH}-$M$_*$ relation \citep[e.g.,][]{Jahnke2009, Merloni2010, Schramm2013, Sun2015, Suh2020, Setoguchi2021, Poitevineau2023} as opposed to the M$_{BH}-$M$_{bulge}$ that is often studied at $\rm z<1$ \citep[e.g.][]{Park2015}.

For the majority of the aforementioned observational studies, the M$_{BH}-$M$_*$ relation has been examined for broad-line AGN  whose M$_{BH}$ were measured using continuum luminosities and broad-line widths. \cite{Jahnke2009} used 10 AGN in the COSMOS field, at $\rm 1<z<2$ and found no difference between their M$_{BH}-$M$_*$ relation and that in the local universe. \cite{Merloni2010} used 89 broad line AGN in the zCOSMOS survey at $\rm 1<1<2.2$ and found that the M$_{BH}/$M$_*$ ratio evolves with redshift. \cite{Schramm2013} used 18 X-ray selected, broad-line AGN at $\rm 0.5<z<1.2$ and found that bulge-dominated host galaxies are more aligned with the local relation than those with prominent disks. \cite{Sun2015} used 69 {\it{Herschel}} detected, broad line AGN at $0.2\leq z <2.1$ and found that galaxies with overmassive (undermassive) black holes, BHs, tend to have a low (high) ratio of the specific accretion rate to the specific star formation rate. \cite{Suh2020} used a sample of 100 X-ray selected, broad-line and moderate luminosity AGN in the {\it{Chandra}}-COSMOS Legacy survey up to $\rm z\sim 2.5$ and found no significant evolution of the M$_{BH}/$M$_*$ ratio. \cite{Setoguchi2021} used 117, moderate luminosity, broad loine AGN in the Subaru/XMM-Newton Deep Field (SXDF) and found that the M$_{BH}/$M$_*$ ratio is similar to that in the local Universe. According to the authors, if their galaxies are bulge dominant, then they have already established the local M$_{BH}-$M$_*$ relation. If they are disk dominant, then their SMBHs are overmassive relative to their M$_*$. 

In most of the above works a limiting factor is the relative small number of AGN ($\lessapprox  100$) used in the analysis. Furthermore, the examination of the evolution of the M$_{BH}-$M$_*$ relation with redshift is done my comparing and combining results from different studies. This approach, though, may hint at systematic effects. Although, the calculation of M$_{BH}$ at different redshifts and therefore using widths of different broad lines (H\,$\beta$, Mg~{\sc ii}, C~{\sc iv}) gives consistent results \citep[e.g.,][]{Shen2013, Liu2016}, this is not true for the measurement of the galaxy properties, such as the M$_*$ and the star-formation rate (SFR). In this case, utilizing different methods and/or different templates and parameter space (e.g., when fitting their spectral energy distribution, SED) may introduce a number of systematics that could affect the comparison and thus the overall conclusions \citep{Mountrichas2021c}.

In this work, we use 687 X-ray selected, broad-line and luminous (median $\rm log\,[L_{X,2-10keV}(ergs^{-1})]=44.3$) AGN, that span a redshift range of $\rm 0.2<z<4.0$ to study the correlation between the AGN and their host galaxy properties, at different redshift intervals. In Sect. 2 we describe the parent sample and the strict criteria we apply to compile a final dataset with accurate and consistent AGN and galaxy measurements. The results are presented in Sect. 3 and we summarize our main conclusions in Sect. 4. 

Throughout this work, we assume a flat $\Lambda$CDM cosmology with $H_ 0=70.4$\,km\,s$^{-1}$\,Mpc$^{-1}$ and $\Omega _ M=0.272$ \citep{Komatsu2011}.

\section{Data}

In this section, we describe the XMM-{\it{XXL}} survey and how we obtained measurements for important AGN and host galaxy properties that are used throughout this work.


\subsection{The sample}

The X-ray AGN used in this study were observed in the North field of the XMM-Newton-XXL survey \citep[XMM-{\it{XXL}};][]{Pierre2016}. XMM-{\it{XXL}} is a medium-depth X-ray survey that covers a total area of 50\,deg$^2$ split into two fields equal in size, the XMM-{\it{XXL}} North (XXL-N) and the XMM-{\it{XXL}} South (XXL-S). The {\it{XXL}}-N sample consists of 8445 X-ray sources. Of these X-ray sources, 5294 have SDSS counterparts and 2512 have reliable spectroscopy \citep{Menzel2016, Liu2016}. Mid-IR and near-IR was obtained following the likelihood ratio method \citep{Sutherland_and_Saunders1992} as implemented in \cite{Georgakakis_Nandra2011}. For more details on the reduction of the {\it{XMM}} observations and the IR identifications of the X-ray sources, see \cite{Georgakakis2017b}. 


\begin{figure}
\centering
  \includegraphics[width=0.75\columnwidth, height=5.8cm]{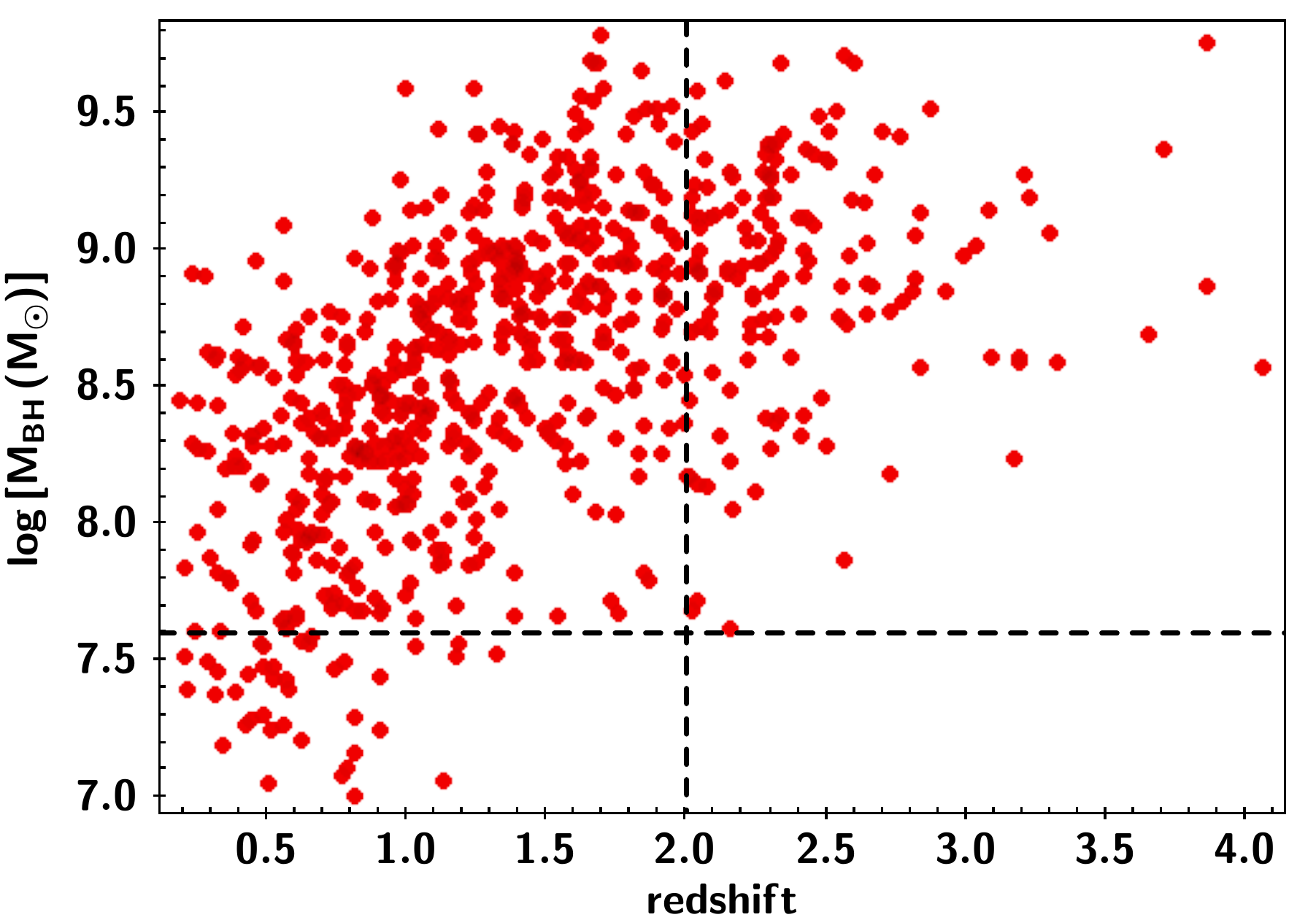} 
  \includegraphics[width=0.75\columnwidth, height=5.8cm]{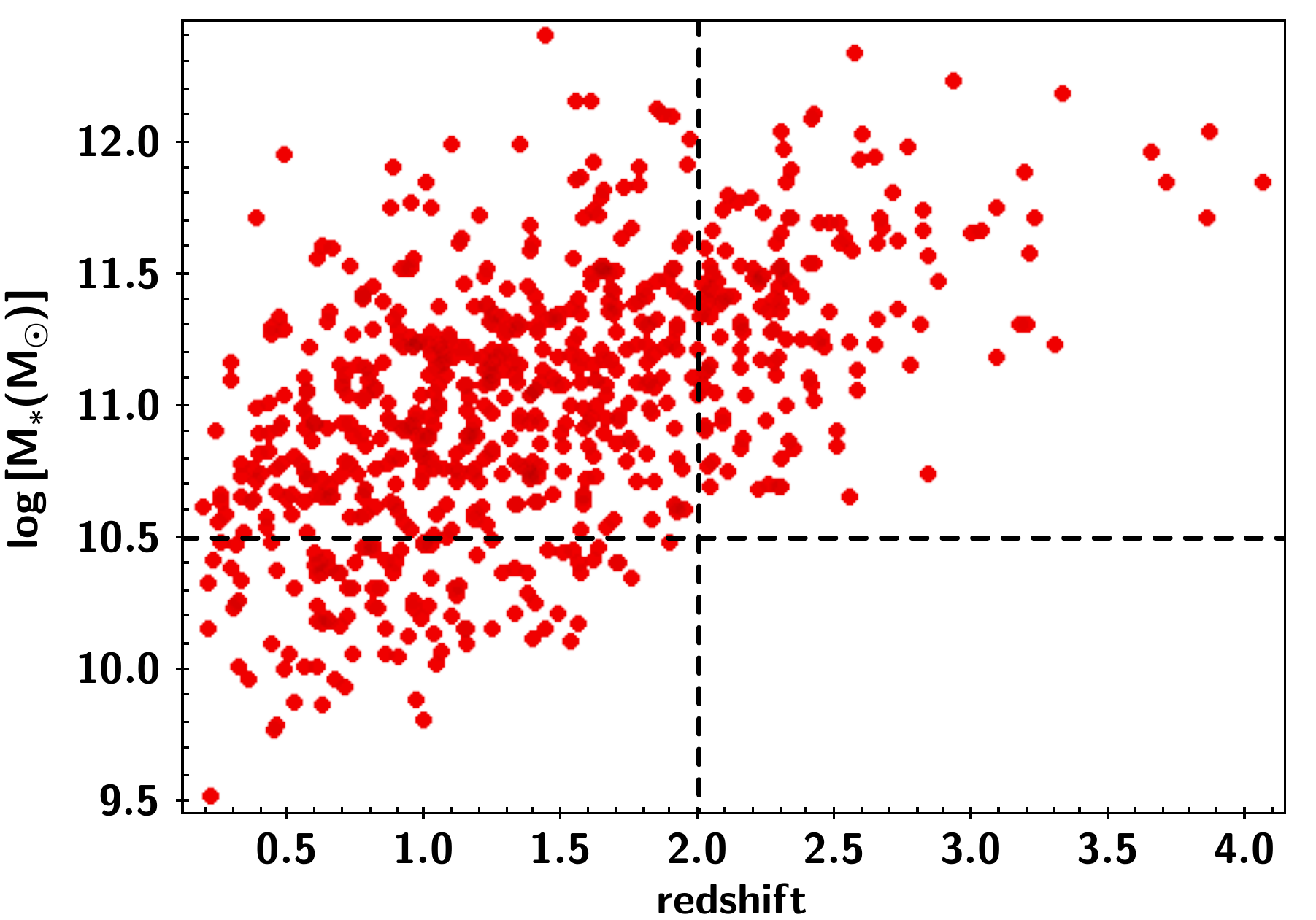} 
  \caption{SMBH and galaxy properties as a function of redshift. Top panel: M$_{BH}$ vs. redshift. Bottom panel: M$_*$ vs. redshift. The dashed horizontal lines show the M$_{BH}$ (top panel) and M$_*$ (bottom panel) limits that our sample is complete up to redshift 2 (vertical lines). For more details, see Sect. 3.3.}
  \label{fig_bias}
\end{figure} 

\subsection{Black hole mass measurements}

As mentioned above, there are 2512 AGN in the XXL-N catalogue that have reliable spectroscopy from SDSS-III/BOSS. 1786 out of these 2512 sources, have been classified as broad line AGN (BLAGN1), by \cite{Menzel2016}. A source was classified as BLAGN1 using the full width at half-maximum (FWHM) threshold of 1000\,Km\,s$^{-1}$. \cite{Liu2016} performed spectral fits to the BOSS spectroscopy of these 1786 BLAGN1 to estimate single-epoch virial M$_{BH}$ from continuum luminosities and broad line widths \citep[e.g.,][]{Shen2013}. The details of the spectral fitting procedure are given in Sect. 3.3 of \cite{Liu2016} and in \cite{Shen2013}. In brief, they first measured the continuum luminosities and broad line FWHMs. Then, they used several single-epoch virial mass estimators to calculate M$_{BH}$. Specifically, they applied the following fiducial mass recipes, depending on the redshift of the source: H\,$\beta$ at $\rm z<0.9$, Mg~{\sc ii} at $\rm 0.9<z<2.2$ and   C~{\sc iv} at $\rm z>2.2$. 

Previous studies have shown that single-epoch M$_{BH}$ estimates that use different emission lines, when adopting the fiducial single-epoch mass formula, are generally consistent with each other with negligible systematic offsets and scatter \citep[e.g.,][]{Shen2008, Shen2011, Shen2012, Shen2013}. \cite{Liu2016} confirmed these previous findings. Finally, their M$_{BH}$ measurements have, on average, errors of $\sim 0.5$\,dex, whereas sources with higher SNR have uncertainties of the measured M$_{BH}$ that are less than 0.15\,dex. 

\subsection{Host galaxy measurements}

In our analysis, we use the host galaxy measurements presented in \cite{Mountrichas2023}. These have been derived by applying spectral energy distribution (SED) fitting, using the CIGALE code \citep{Boquien2019, Yang2020, Yang2022}. The available photometry has been compiled and presented in \cite{Masoura2018, Masoura2021}. The templates and parameter space used is the same as that presented in \cite{Mountrichas2021c, Mountrichas2022a, Mountrichas2022b}. In brief, a delayed star formation history (SFH) model with a function form $\rm SFR\propto t \times exp(-t/\tau$) is used to fit the galaxy component. A star formation burst is included \citep{Ciesla2017, Malek2018, Buat2019} as a constant ongoing period of star formation of 50\,Myr. The \cite{Bruzual_charlot2003} single stellar population template is used to model the stellar emission. Stellar emission is attenuated following \cite{Charlot_Fall_2000}. The dust heated by stars is modelled following \cite{Dale2014}. The SKIRTOR template \citep{Stalevski2012, Stalevski2016} is used for the AGN emission. Accounting for the AGN emission significantly reduces the biases on the estimate of the M$_*$ and SFR of their host galaxy \citep{Ciesla2015}. The values for the various parameters are similar to those presented in Tables 1 in \cite{Mountrichas2021c, Mountrichas2022a, Mountrichas2022b}. 

To examine if our SFR and M$_*$ measurements are sensitive to the adopted SFH model, we repeat the SED fitting process using a different SFH module. Specifically, we adopt an expansion of the delayed SFH model mentioned above, that allows for a recent quenching of the SFR \citep{Ciesla2017}. This module is provided as \texttt{sfhdelayedbq} in CIGALE. We confirm that CIGALE measurements for the host galaxy properties of interest are robust and do not depend on the selection of the SFH model \citep[see also Appendix B in][]{Mountrichas2022b}.


\subsection{Final sample}

To ensure that only sources with reliable galaxy measurements are included in our analysis, we follow the criteria used in \cite{Mountrichas2021c, Mountrichas2022a, Mountrichas2022b, Mountrichas2023}. Specifically, we include only sources that have measurements in the following photometric bands: $u$, $g$, $r$, $i$, $z$, J, H, K, W1, W2 and W4, where W1, W2 and W4 are the WISE photometric bands at 3.4, 4.6 and 22\,$\mu$m. Approximately, $50\%$ of our sources have far-IR measurements by Herschel \citep[HELP collaboration;][]{Shirley2019, Shirley2021}. However, previous studies have shown that lack of far-IR photometry does not affect the SFR calculations of CIGALE \citep{Mountrichas2021a, Mountrichas2021c, Mountrichas2022b, Mountrichas2022a}. Therefore, we do not require our sources to have available far-IR photometry.

Furthemore, we exclude sources with bad SED fits and unreliable host galaxy measurements. Towards this end, we impose a  reduced $\chi ^2$  threshold of $\chi ^2_r <5$ \citep[e.g.][]{Masoura2018, Buat2021}. We also exclude systems for which CIGALE could not constrain the parameters of interest (SFR, M$_*$). For that we apply the same criteria used in previous recent studies \citep[e.g.][]{Mountrichas2021c, Buat2021, Mountrichas2022a, Mountrichas2022c, Koutoulidis2022}. The method uses the two values that CIGALE provides for each estimated galaxy property. One value corresponds to the best model and the other value (bayes) is the likelihood-weighted mean value. A large difference between the two calculations suggests a complex likelihood distribution and important uncertainties. We therefore only include in our analysis sources with $\rm \frac{1}{5}\leq \frac{SFR_{best}}{SFR_{bayes}} \leq 5$ and $\rm \frac{1}{5}\leq \frac{M_{*, best}}{M_{*, bayes}} \leq 5$, where SFR$\rm _{best}$ and  M$\rm _{*, best}$ are the best-fit values of SFR and M$_*$, respectively and SFR$\rm _{bayes}$ and M$\rm _{*, bayes}$ are the Bayesian values estimated by CIGALE. These criteria reduce the number of X-ray AGN to 1592. We note that this number includes sources with either spectroscopic or photometric redshifts (photoz). The photoz calculations have been presented in \cite{Masoura2018}. 

We then cross match these 1592 AGN with the spectroscopic sample of \cite{Liu2016}. This results in 687 broad-line, X-ray AGN with spectroscopic redshifts that have reliable measurements for M$_{BH}$ and galaxy properties. The distribution of M$_{BH}$ and M$_*$ vs. redshift for our final AGN sample is presented in the top and bottom panels of Fig. \ref{fig_bias}, respectively. The X-ray luminosity of the sources spans a range of $\rm 42.5<log\,[L_{X,2-10keV}(ergs^{-1})]<45.5$ with a median value of $\rm log\,[L_{X,2-10keV}(ergs^{-1})]=44.3$. The median redshift is $\rm z=1.4$ ($\rm 0.2<z<4.0$). To minimize selection effects in our analysis (see next sections), we split the dataset into four redshift intervals. There are 181 AGN at $\rm z<0.9$, 215 at $\rm 0.9<z<1.5$, 188 at $\rm 1.5<z<2.2$ and 103 AGN at $\rm z>2.2$. 

There are two available measurements for the bolometric luminosities, L$_{bol}$, of our sources. The catalogue of \cite{Liu2016} includes L$_{bol}$ calculations. For their estimation, they have integrated the radiation directly produced by the accretion process, that is the thermal emission from the accretion disc and the hard X-ray radiation produced by inverse-Compton scattering of the soft disc photons by a hot corona (for more details see their Sect. 4.2). We also combine CIGALE's measurements for the absorption corrected X-ray luminosity and the AGN disc luminosity. Comparison of the two L$_{bol}$ estimates shows that the distribution of their ($\rm log$) difference has a mean value of 0.08\,dex and a standard deviation of 0.42\,dex. In our analysis, we choose to use the L$_{bol}$ calculations of CIGALE. However, we confirm that this choice does not affect our overall results and conclusions.

\section{Results and Discussion}

\begin{figure}
\centering
  \includegraphics[width=0.9\columnwidth, height=6.5cm]{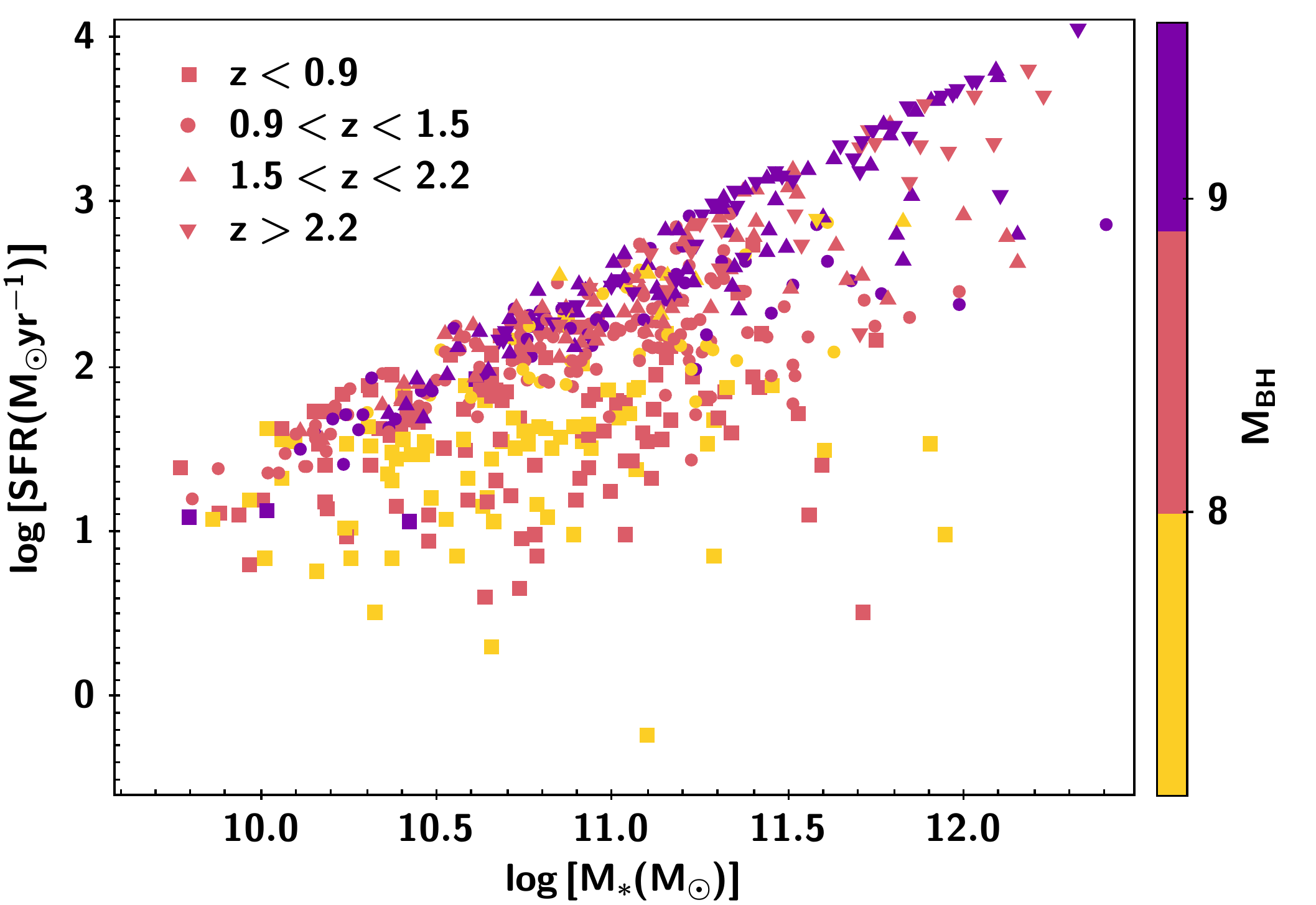} 
  \caption{SFR vs. M$_*$ of the X-ray AGN in our dataset. Different symbols indicate different redshift intervals, as shown in the legend. The results are colour-coded based on the M$_{BH}$ value.}
  \label{fig_ms}
\end{figure}

\subsection{SFR vs. M$_*$}
\label{sec_ms}

The SFR-M$_*$ relation of our sources is presented in Fig. \ref{fig_ms}. Different symbols correspond to different redshift intervals, as indicated in the legend. The results are colour-coded based on the M$_{BH}$. Sources located in the upper, right corner of the SFR-M$_*$ space, with $\rm log\,[SFR (M_\odot yr^{-1})]>3$ and $\rm log\,[M_*(M_\odot)]>12$ are high redshift sources ($\rm z>2$) with massive SMBHs ($\rm log\,[M_{BH}\,(M_\odot)]>9$). Based on the sSFR ($\rm sSFR=SFR/M_*$) measurements of CIGALE, there is only a handful of AGN in our dataset that are in quiescent systems ($\rm log\,sSFR(Gyr^{-1})>11$). The vast majority of our AGN are either in star-forming or in starburst galaxies. We note that we chose to identify quiescent systems based on their sSFR values as opposed to overploting a MS from the literature \citep[e.g.][]{Whitaker2014, Schreiber2015} due to the systematics that this approach may introduce \citep[for more details see e.g.,][]{Mountrichas2021c}.  

Most previous studies have found a strong, positive correlation between the SFR and the X-ray luminosity of AGN \citep[e.g.,][]{Lanzuisi2017, Masoura2018, Masoura2021}, although more recent works have shown that this correlation is weaker when systematic effects are minimized and the M$_*$ is taken into account \citep[e.g.,][]{Mountrichas2021c,Mountrichas2022a, Mountrichas2022b}. Recently, \cite{Bluck2023} analyzed three cosmological hydrodynamical simulations (Eagle, Illustris and IllustrisTNG) and concluded that the M$_{BH}$ is the predictive parameter of galaxy quenching and not the AGN luminosity. We apply Pearson correlation analysis splitting our data into the four redshift intervals, shown in the legend of Fig. \ref{fig_ms}. The results show that the sSFR and X-ray luminosity (or L$_{bol}$) have an (average over the four redshift bins) correlation coefficient of $\rm r=0.45\pm 0.15$ (or $\rm r=0.37\pm 0.22$). The correlation coefficient of the sSFR and M$_{BH}$ is $\rm r=0.31\pm 0.12$. The errors are the standard deviations. These results, although indicative, corroborate that observational works should examine the role of M$_{BH}$ when studying the impact of AGN feedback on the host galaxy properties, as suggested by \cite{Bluck2023}. 

\begin{figure}
\centering
  \includegraphics[width=0.9\columnwidth, height=7cm]{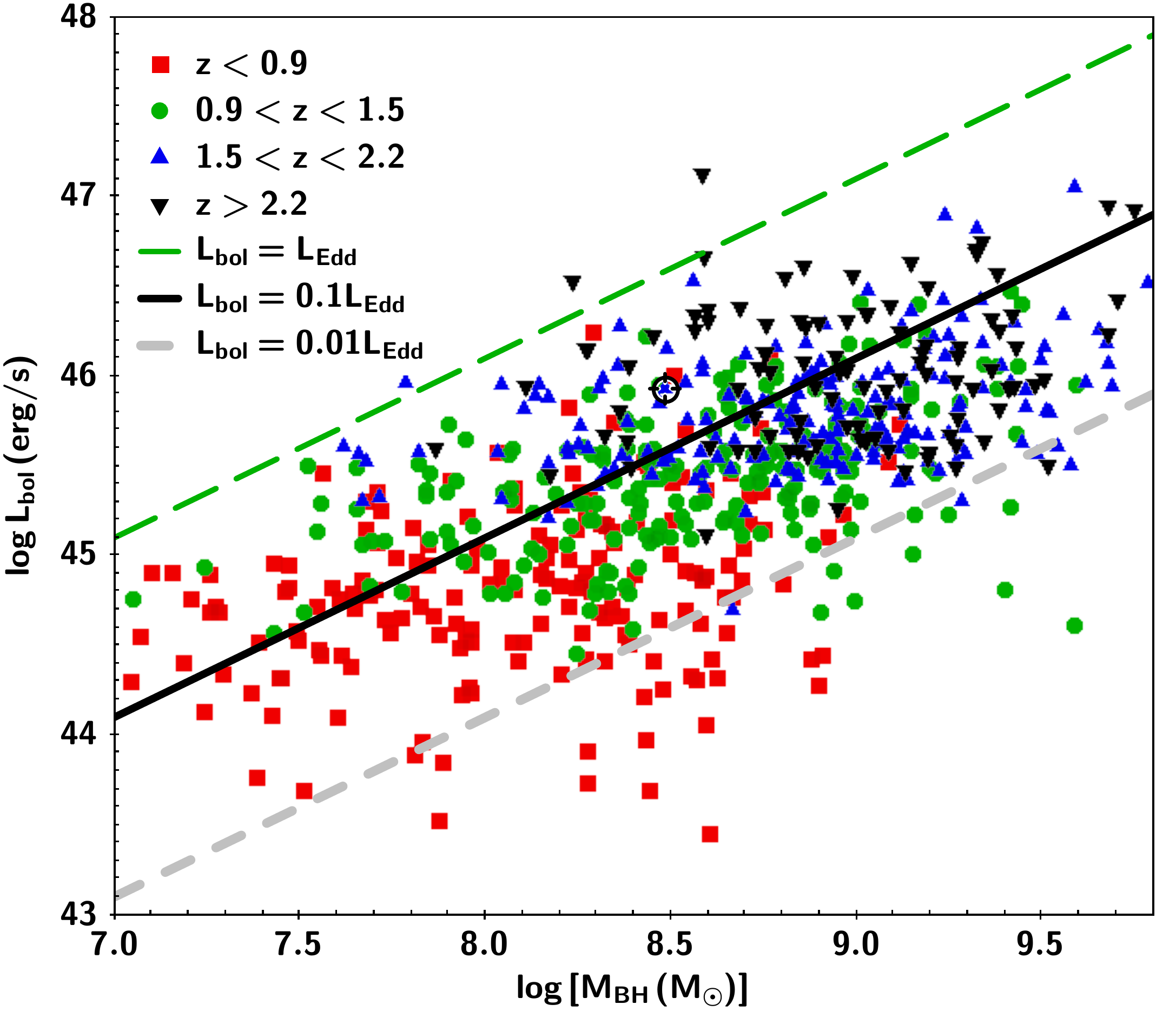} 
  \caption{L$_{bol}$ as a function of M$_{BH}$. Different symbols and colours correspond to different redshift intervals, as indicated in the legend. The lines correspond to L$_{bol}=$L$_{Edd}$ (green, long-dashed line), L$_{bol}=0.1$\,L$_{Edd}$ (solid, black line) and L$_{bol}=0.01\,$L$_{Edd}$ (grey, short-dashed line).}
  \label{fig_lbol_vs_mbh}
\end{figure}

\subsection{L$_{bol}$ vs. M$_{BH}$}
\label{sec_lbol_mbh}

In Fig. \ref{fig_lbol_vs_mbh}, we plot the L$_{bol}$ calculations of CIGALE as a function of M$_{BH}$ for the different redshift intervals used in our study. The lines correspond to L$_{bol}=$L$_{Edd}$ (green, long-dashed line), L$_{bol}=0.1$\,L$_{Edd}$ (solid, black line) and L$_{bol}=0.01\,$L$_{Edd}$ (grey, dashed line), where L$_{Edd}=1.26 \times 10^{38}\rm M_{BH}/M_\odot\,erg\,s^{-1}$. The vast majority of our AGN lie between Eddington ratios of 0.01 to 1, with a median value of 0.06, in agreement with previous studies \citep[e.g.,][]{Trump2009, Lusso2012, Sun2015, Suh2020}. The fact that the SMBHs of the X-ray luminous AGN in our dataset accrete at sub-Eddington rates, at all redshifts spanned by our sample, could indicate that most of their mass has been built up at earlier epochs. 

\begin{figure}
\centering
  \includegraphics[width=0.9\columnwidth, height=7cm]{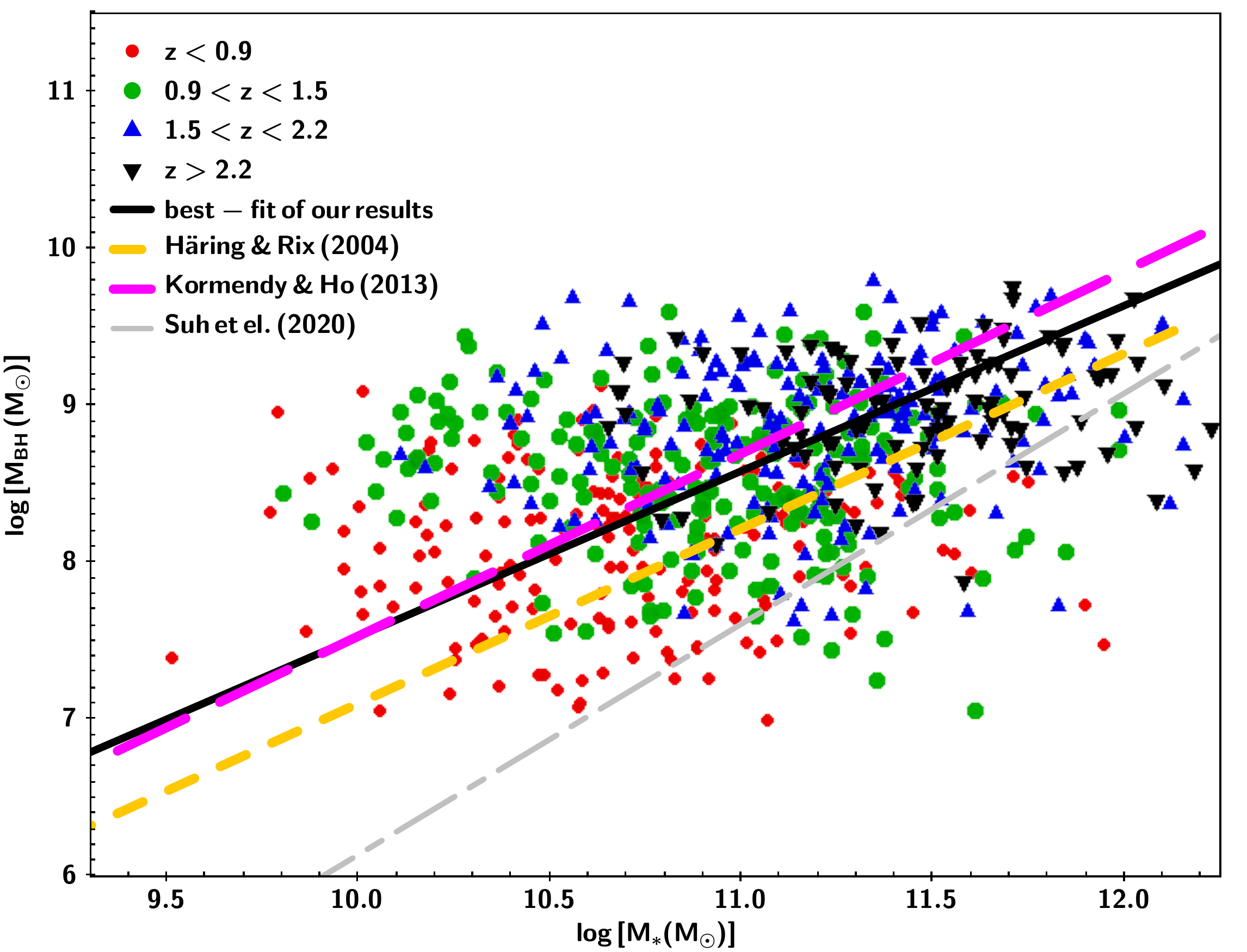} 
  \includegraphics[width=0.9\columnwidth, height=7cm]{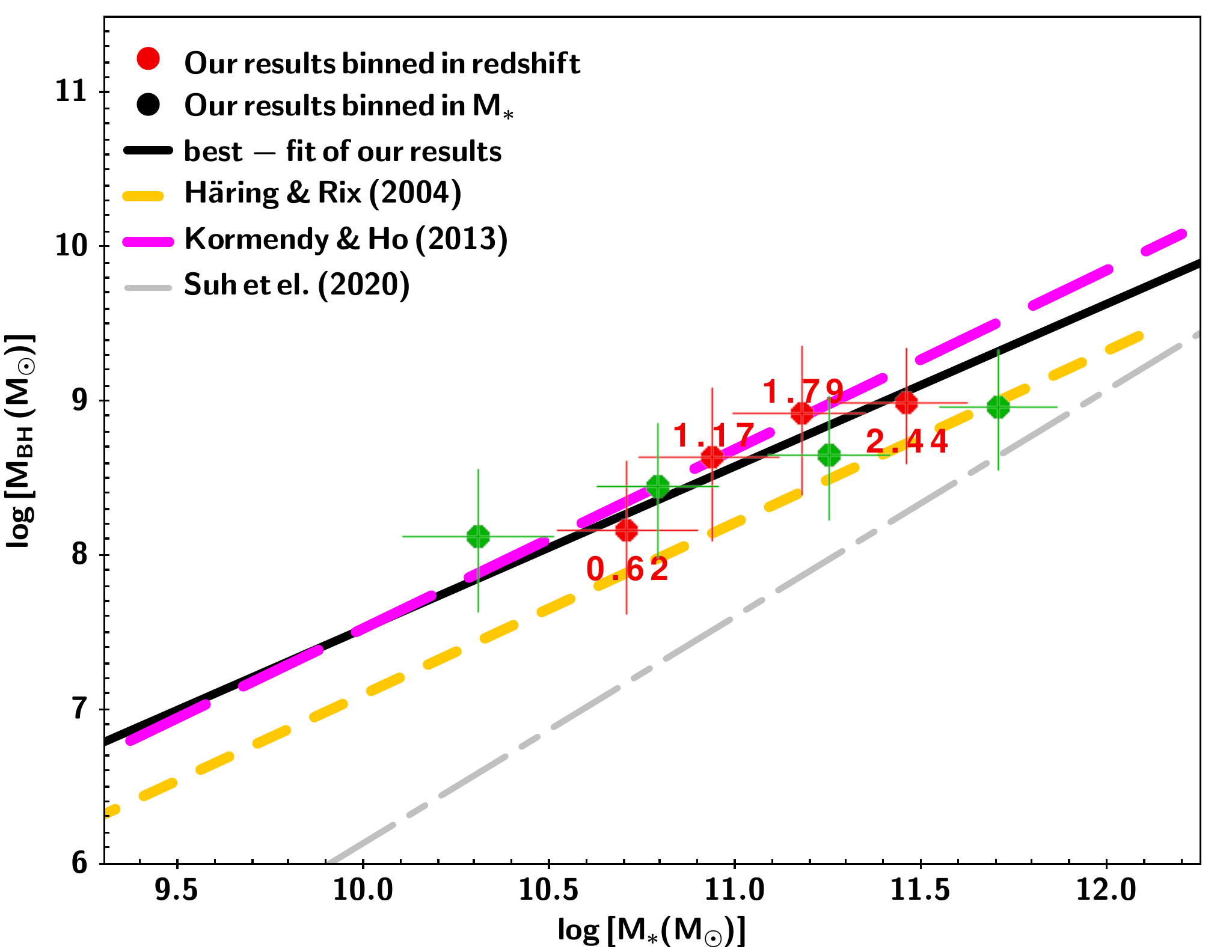} 
  \caption{M$_{BH}$ as a function of M$_*$. Top panel shows the distribution of our AGN in the M$_{BH}-$M$_*$ space. Different colours correspond to different redshift bins. The bottom panel shows our results grouped in bins of redshift (red symbols) and M$_*$ (green symbols). Median values are presented as well as the $1\,\sigma$ uncertainties (calculated via bootstrap resampling). The labels next to the red points show the median redshift that corresponds to each bin. The solid, black line shows the best-fit on our measurements. The dashed lines, in both panels, present the best-fits from previous studies (see text for more details).  }
  \label{fig_mbh_vs_mstar}
\end{figure} 

\begin{figure}
\centering
  \includegraphics[width=0.75\columnwidth, height=5.8cm]{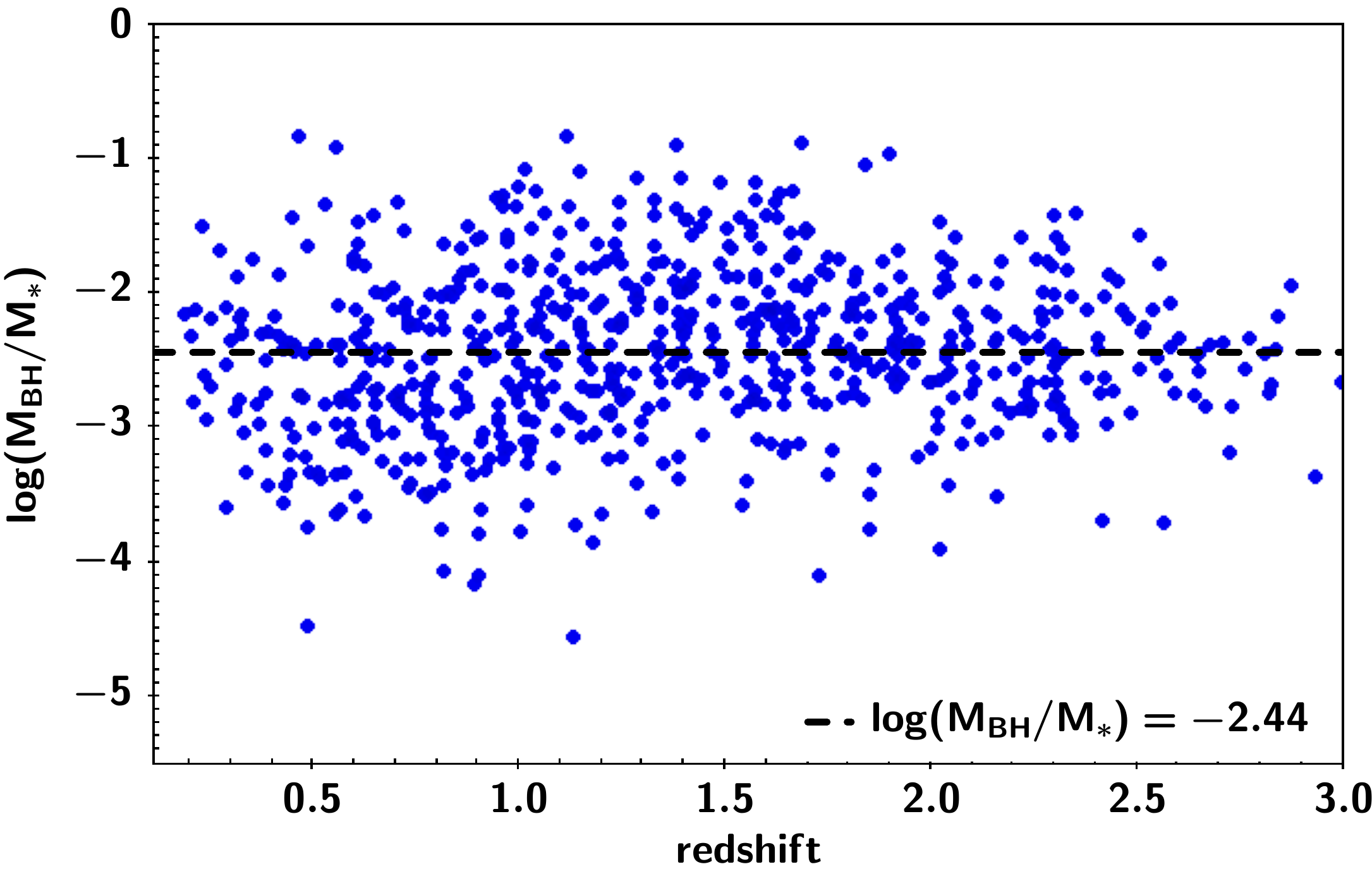} 
  \includegraphics[width=0.75\columnwidth, height=5.8cm]{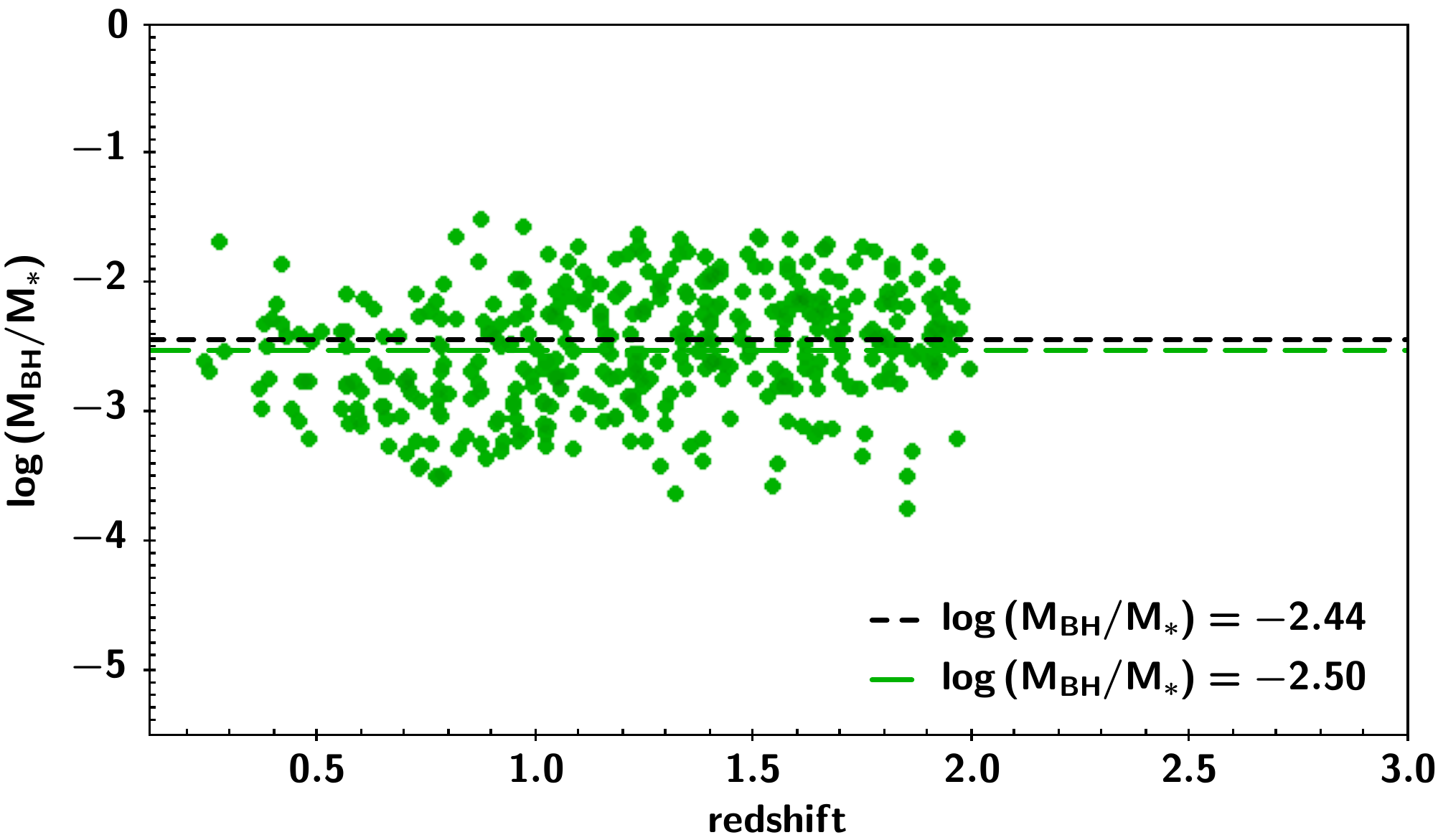} 
  \caption{M$_{BH}-$M$_*$ ratio as a function of redshift. Top panel shows the values for the total AGN sample. Bottom panels shows the results when we account for selection biases (see text for more details). The dashed, black, horizontal line, in both panels, indicates the mean M$_{BH}-$M$_*$ ratio value of our dataset. The green, long-dashed, horizontal line presents the mean M$_{BH}-$M$_*$ ratio value, when selection biases are taken into consideration.}
  \label{fig_mbh_mstar_vs_redz}
\end{figure}

\subsection{The M$_{BH}-$M$_{*}$ relation}
\label{sec_mstar_mbh}

In this section, we examine the M$_{BH}-$M$_{*}$ relation and the evolution of the M$_{BH}/$M$_*$ ratio with redshift. The top panel of Fig. \ref{fig_mbh_vs_mstar} presents the M$_{BH}$ measurements of the AGN as a function of their host galaxy M$_*$. Symbols are colour coded based on the redshift interval that the source belongs. The solid, black line presents the best-fit of our measurements (M$_{BH}=1.054$\,M$_*-3.010$). In the same panel, we also present the best fits from previous studies. The orange, dashed line shows the best fit from \cite{Haring2004}. In that study, the authors examined the black hole$-$bulge mass relation, using a sample of 30 nearby galaxies. The purple, long-dashed line, presents the best fit from \cite{Kormendy2013} for local, bulge galaxies. The grey, dashed line is the best fit found by \cite{Suh2020}, that combined their sample of 100 X-ray selected AGN in the $\it{Chandra}-$COSMOS Legacy Survey that spanned a redshift range up to 2.5, with the sample presented in \cite{Reines2015} that consists of nearby, inactive early-type galaxies as well as local AGN. 

We apply a correlation analysis and we find a correlation coefficient of $r=0.47\pm 0.21$ between the two properties for the sources in our sample. This value is averaged over the four redshift intervals. The error presents the standard deviation of the four measurements. This result is consistent with the value we calculate using the M$_*$ and M$_{BH}$ measurements presented in Table 2 in \cite{Sun2015} that have a median redshift of $\approx 1.2$ ($r=0.43$). It is also consistent with  M$_{BH}-$M$_{*}$ correlations reported in the local Universe \citep[e.g., 0.54;][]{Reines2015}. The correlation coefficients of our M$_{BH}$ and M$_*$ for each of the redshift bins are: 0.31, 0.35, 0.38, 0.36, at $\rm z<0.9$, $\rm 0.9<z<1.5$, $\rm 1.5<z<2.2$ and $\rm z>2.2$, respectively. Although, the coefficient values are consistent across all redshifts spanned by our dataset, they appear a bit lower compared to the correlation coefficient of the full sample. This is, probably, mostly due to the smaller M$_{BH}$ ranges spanned by the individual redshift bins compared to the full catalogue. The bottom panel of Fig. \ref{fig_mbh_vs_mstar} presents the results when we bin the M$_{BH}$ and M$_*$ into the four redshift bins, we use in our analysis (red circles). The median M$_{BH}$ and M$_*$ values and their 1\,$\sigma$ uncertainties (calculated via bootstrap resampling) are shown. We also bin our M$_*$ and M$_{BH}$ calculations in four M$_*$ bins within $10<\rm log\,[M_*(M_\odot)]<12$ (green circles). Each bin has 0.5\,dex width. 


Next, we examine the M$_{BH}/$M$_*$ ratio as a function of redshift. Based on the results presented in the top panel of Figure \ref{fig_mbh_mstar_vs_redz}, the M$_{BH}/$M$_*$ ratio does not evolve with redshift. The mean $\rm log($M$_{BH}/$M$_*)$ value is found at $-2.44$ (with a 1$\sigma$ scatter of 0.61), shown by the dashed line. The scatter of the $\rm log($M$_{BH}/$M$_*)$ ratio is similar at all redshifts spanned by the dataset. Specifically, $\sigma \sim 0.6$ up to redshift 2.2 and $\sigma \sim 0.45$ at $\rm z>2.2$. The value of the $\rm log($M$_{BH}/$M$_*)$ ratio is in agreement with that found by \cite{Setoguchi2021} (-2.2), but somewhat higher compared to that found by \cite{Suh2020} ($\approx -2.7$) and that found in the local universe \citep[-2.85;][]{Haring2004}. 

We note that our dataset is a high redshift, flux limited sample and this make it susceptible to suffer from Eddington bias. However, as has pointed out, for instance in \cite{Poitevineau2023}, both M$_{BH}$ and M$_*$ scale with with BLR line and the infrared-to-optical luminosities, respectively, and thus have a similar dependence on redshift. Furthermore, the M$_{BH}/$M$_*$ ratio spans a wide range, at least up to redshift 2, as shown in the top panel of Fig. \ref{fig_mbh_mstar_vs_redz}. We also note that the mean value of the $\rm log($M$_{BH}/$M$_*)$ ratio of the full sample is similar to that found in each of the redshift bins used in our study. Specifically, we find $\rm log($M$_{BH}/$M$_*)=-2.52, -2.34, -2.29, -2.49$ at $\rm z<0.9$, $\rm 0.9<z<1.5$, $\rm 1.5<z<2.2$ and $\rm z>2.2$, respectively.

Nevertheless, to minimize possible selection biases, we choose from our dataset sources that lie in a M$_*$ and M$_{BH}$ space that is detected throughout $\rm z=2$. Specifically, we choose sources that fulfil the following criteria: $\rm z\leq2$,  $\rm log\,[M_*(M_\odot)]>10.5$ and $\rm log\,[M_{BH}\,(M_\odot)]>7.6$ (dashed lines in Fig. \ref{fig_bias}). 423 AGN meet these requirements. Their M$_{BH}/$M$_*$ ratio as a function of the redshift is presented in the bottom panel of Fig. \ref{fig_mbh_mstar_vs_redz}. No statistical significant evolution of the M$_{BH}/$M$_*$ ratio with redshift is detected. The value of the M$_{BH}/$M$_*$ ratio is similar in the four redshift bins used in our analysis, with a mean value of $-2.50$ (shown by the green, dashed line), which is close to the value found for the total sample. This confirms our previous finding of no evolution of the M$_{BH}/$M$_*$ ratio with redshift. 

We conclude that the M$_{BH}/$M$_*$ ratio does not evolve with cosmic time, at least up to $\rm z=2$, even when we account for selection effects. This is in agreement with most similar studies \citep[e.g.,][]{Jahnke2009, Sun2015, Suh2020, Setoguchi2021} that examined the  M$_*/$M$_{BH}-\rm z$ relation using moderate luminosity X-ay AGN. \cite{Merloni2010} reported an evolution of the M$_{BH}/$M$_*$ ratio with redshift, by comparing their results with those in the local Universe \citep{Haring2004}. However, based on the more recent results of \cite{Kormendy2013} in the local Universe, no evolution would have been detected \citep[see Sect. 3.3 in][]{Setoguchi2021}.  


\subsection{The L$_{bol}$ as a function of SFR }
\label{sec_sfr_lbol}

\begin{figure}
\centering
  \includegraphics[width=0.9\columnwidth, height=6.5cm]{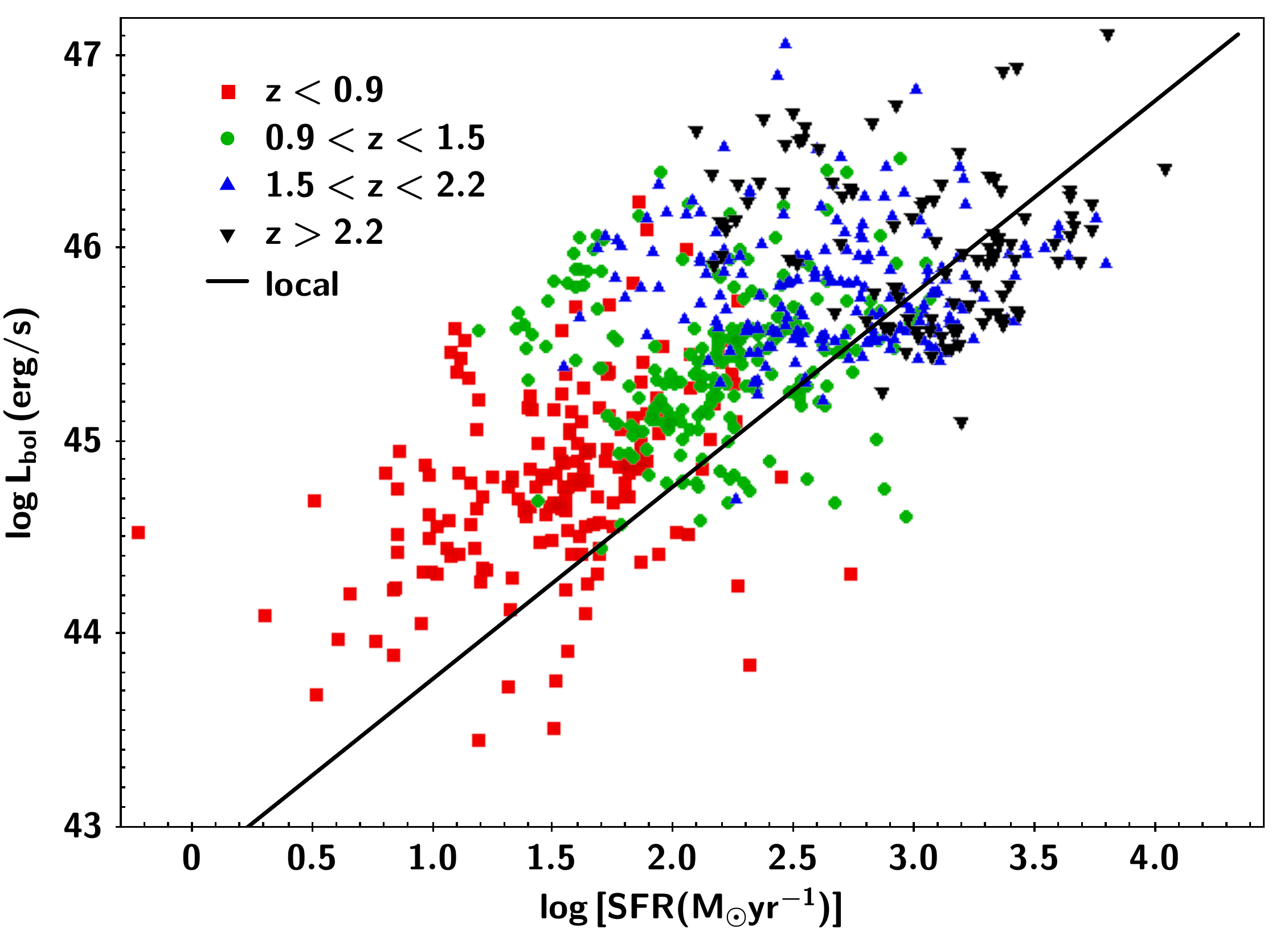} 
  \includegraphics[width=0.9\columnwidth, height=6.5cm]{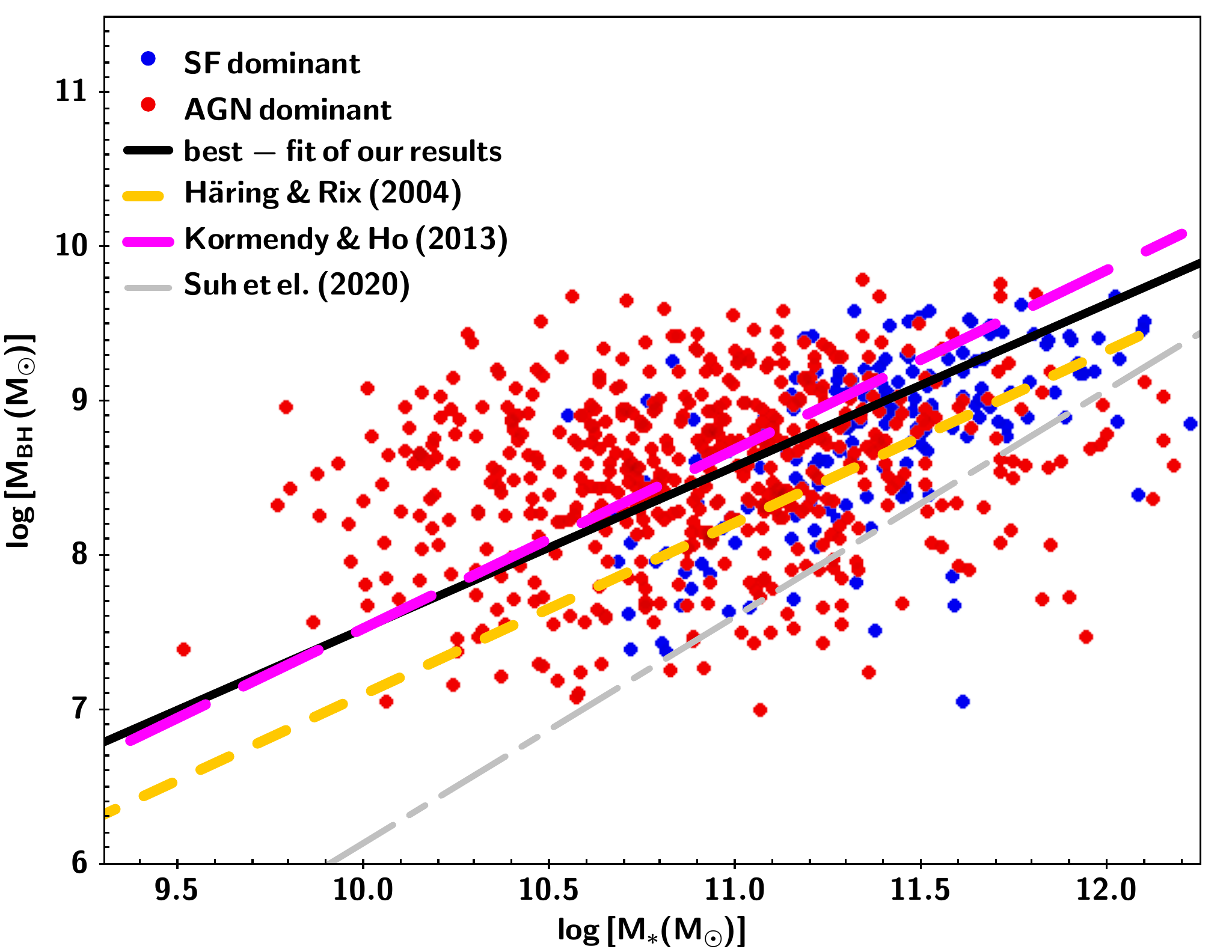} 
  \caption{Correlation of AGN and host galaxy properties. The top panel presents the L$_{bol}$ as a function of SFR. Different symbols and colours correspond to different redshift intervals, as shown in the legend. The solid line indicates the local M$_{BH}-$M$_*$ relation that would be expected from exactly simultaneous evolution of the SMBH and the host galaxy. AGN that lie above this line live in galaxies that the SMBH growth is dominant. AGN that lie below the line are systems in which the galaxy growth dominates. The bottom panel shows the distribution of AGN (red circles) and SF (blue circles) dominant systems in the M$_{BH}-$M$_*$ space. The dashed lines are the same as those used in Fig. \ref{fig_mbh_vs_mstar}.}
  \label{fig_lbol_sfr}
\end{figure}

In the previous section, we examined the relation between M$_*$ and M$_{BH}$. Here, we investigate the relation between their time derivatives, that is the SFR and L$_{bol}$. The results are shown in the top panel of Fig. \ref{fig_lbol_sfr}. To minimize selection effects, the measurements are divided into four redshift bins, as shown in the legend of the figure. The solid line corresponds to the local M$_{BH}-$M$_*$ relation that would be expected from exactly simultaneous evolution of the SMBH and the host galaxy \citep[see Sect. 4.4 and 3.4 in][ respectively]{Ueda2018,Setoguchi2021}. Most of the AGN ($75\%$) lie above the solid line. This implies that these sources are in a SMBH growth phase (AGN dominant systems). However, there is a smaller number of AGN (173) that lie below the solid line, which indicates that in these systems the galaxy growth is dominant (SF dominant systems). 

The systems in which the SF is dominant are mainly high redshift galaxies (median $\rm z=1.85$ compared to $\rm z=1.21$ for the AGN dominant systems) and are in starburst phase ($\rm log\,sSFR(Gyr^{-1})>0.5$). They also have similar M$_{BH}$ with their AGN dominated counterparts, at fixed redshift. SF dominated galaxies also have, on average, lower $\rm log($M$_{BH}/$M$_*)$ ratio ($-2.63$) compared to AGN dominated systems ($-2.39$).

The bottom panel of Fig. \ref{fig_lbol_sfr} presents the M$_{BH}-$M$_*$ relation for the two AGN populations. A pearson correlation analysis yields an average (over the four redshift intervals) of $\rm r=0.62\pm 0.15$ and  $\rm r=0.29\pm 0.11$, for the SF dominated and the AGN dominated galaxies, respectively. This implies a significantly higher correlation between the M$_{BH}$ and M$_*$ for the systems that SF is dominant compared to those that the SMBH growth dominates. 

We repeat the same exercise utilizing the sample of 69 AGN from the COSMOS and CDFS fields, presented in \cite{Sun2015} and using the values shown in their Table 2. We identify six AGN that are in SF dominated systems. The median redshift of the two AGN populations is similar ($\rm z=1.32$ for systems that the galaxy growth is dominant and $\rm z=1.16$ for galaxies that the SMBH growth is dominant). A correlation analysis yields $\rm r=0.41$ and $\rm r=0.86$, for the AGN dominated and SF dominated systems, respectively. The $\rm log($M$_{BH}/$M$_*)$ ratio is $-2.60$ and $-2.99$ for the galaxies that the SMBH growth dominates and for systems that the galaxy growth dominates, respectively. It is worth mentioning that using L$_{bol}$ and SFR values from the literature and compare them with our measurements, may hint at systematics (this is also true when comparing our M$_*$ and M$_{BH}$ with calculations from the literature), as different methods have been applied for the calculation of these parameters \citep[see e.g.,][]{Mountrichas2021c}. Taking into account this caveat and the size of the sample of \cite{Sun2015} (in particular the small number of SF dominated systems it includes), these results confirm the trends we find in our dataset. This could also provide an (alternative) possible explanation for the higher M$_{BH}/$M$_*$ ratio value presented by \cite{Setoguchi2021} and the lack of correlation they found regarding the M$_{BH}-$M$_*$ relation (see the discussion in their Sect. 3.3). Their AGN dataset consists exclusively of AGN dominated galaxies (see their Fig. 4). 

AGN dominated galaxies appear to have, on average, lower M$_*$ compared to their SF dominated counterparts with similar M$_{BH}$ (bottom panel of Fig. \ref{fig_lbol_sfr}). This, in conjunction with their lower correlation between the M$_{BH}$ and M$_*$ compared to SF dominated systems, could suggest that SF growth becomes dominant at a later stage compared to the SMBH growth, moving these galaxies rightwards in the M$_{BH}-$M$_*$ plane and in line with local scaling relations. This could also explain the higher M$_{BH}/$M$_*$ ratio of the AGN dominated systems, in the sense that their M$_*$ growth lacks behind compared to their M$_{BH}$ (or equivalently that their black holes are overweight compared to their stellar mass). 




\begin{figure}
\centering
  \includegraphics[width=0.9\columnwidth, height=6.5cm]{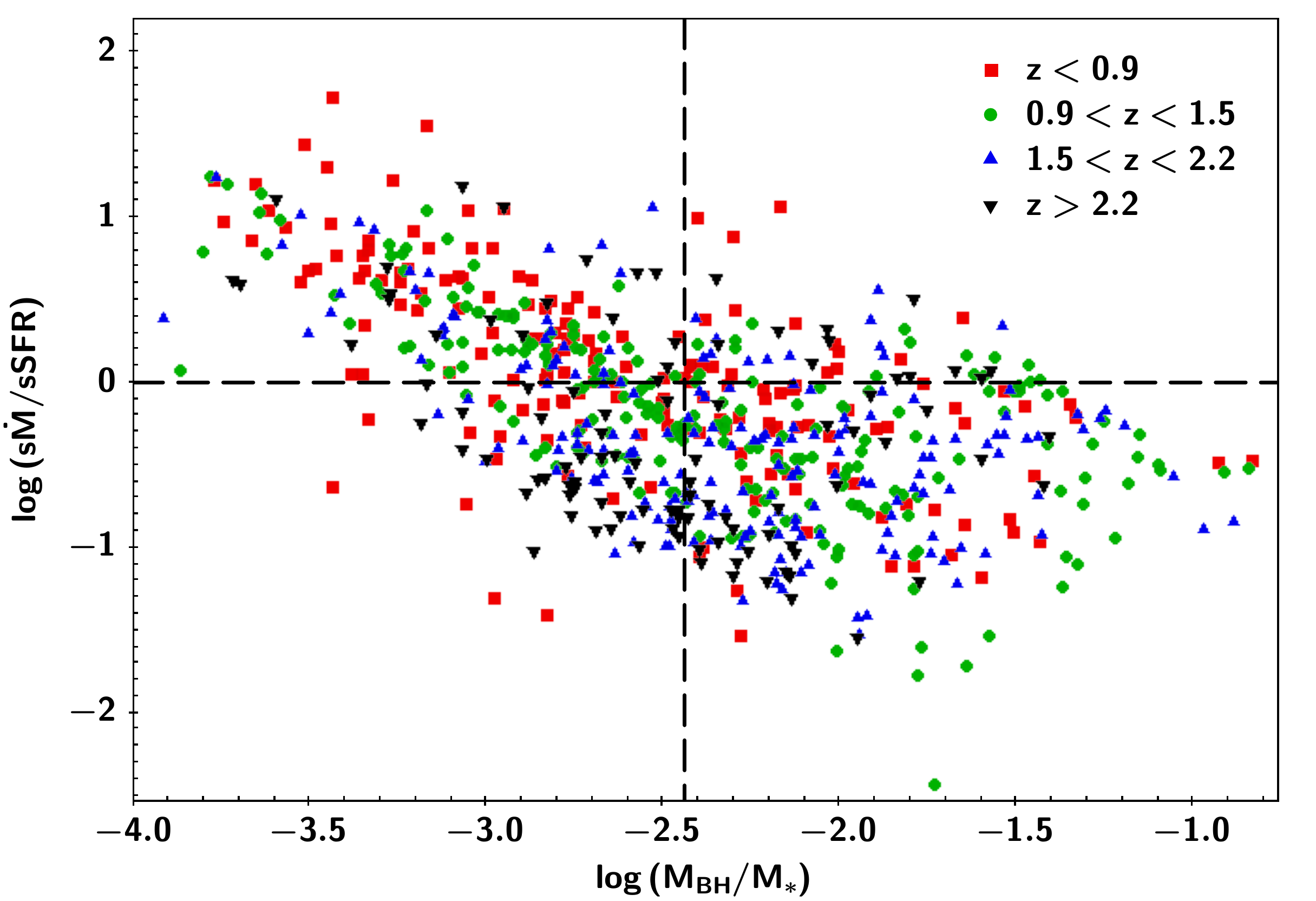} 
  \includegraphics[width=0.9\columnwidth, height=6.5cm]{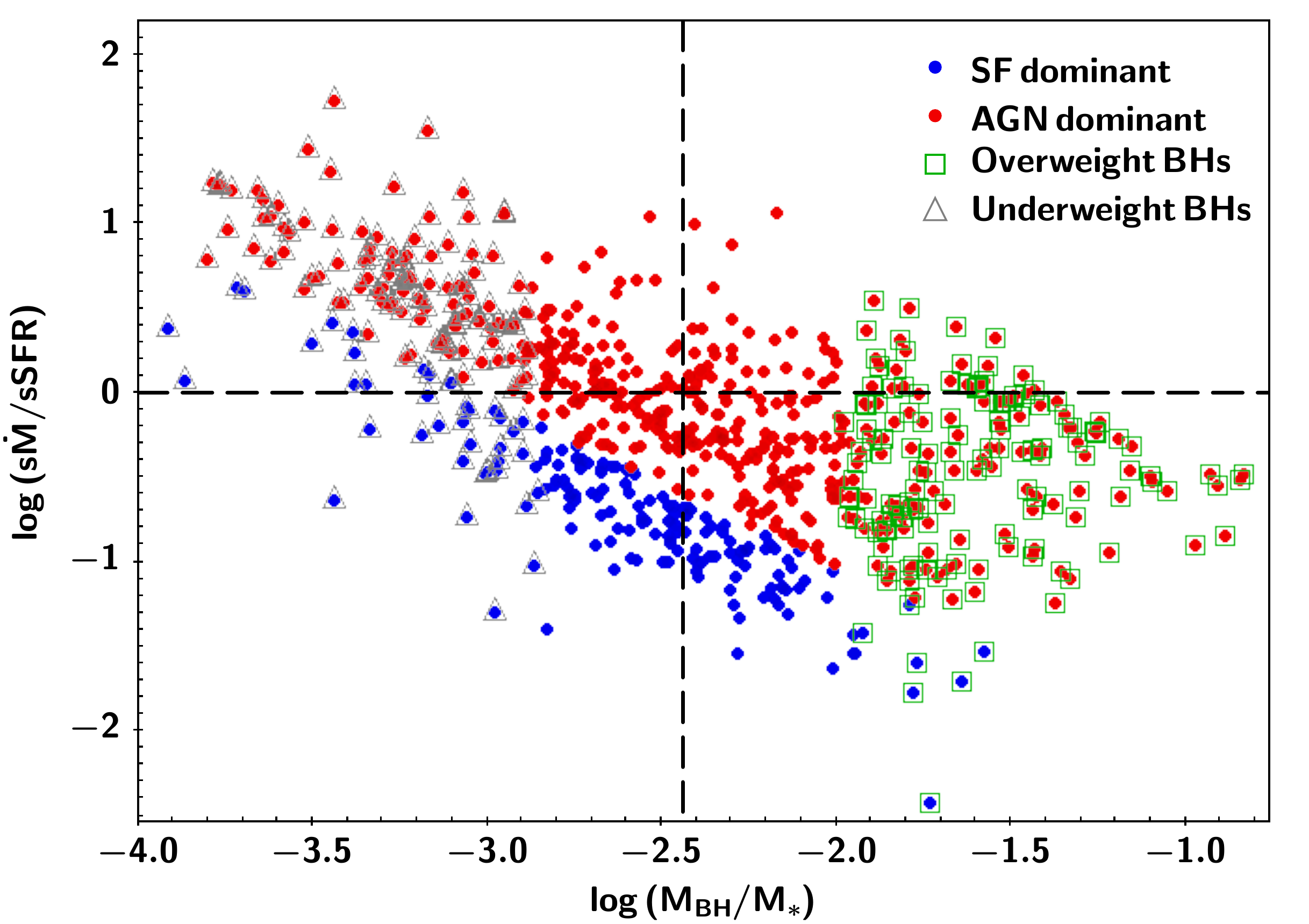} 
  \caption{Distribution of AGN in the $\rm s\dot{M}/$sSFR$-$M$_{BH}/$M$_*$ space. In the top panel, different colours and symbols correspond to different redshift bin, as shown in the legend. The bottom panel presents the distribution of SF (blue circles) and AGN (red circles) in the the $\rm s\dot{M}/$sSFR$-$M$_{BH}/$M$_*$ space. Sources marked with a square are classified as galaxies with overweight SMBHs, while those marked with a triangle are classified as galaxies with underweight SMBHs. The classification is based on the location of the AGN in the M$_{BH}-$M$_*$ plane relative to the line that describes the best-fit on our M$_{BH}-$M$_*$  measurements (see text and Fig. \ref{fig_mbh_vs_mstar} for more details).}
  \label{fig_specific}
\end{figure} 

\subsection{SMBH mass growth rate vs. galaxy stellar mass growth rate}
\label{sec_specific}

To examine the possible pathways of the AGN in the M$_{BH}-$M$_*$ plane, we calculate the specific SMBH growth rate, defined as $\rm s\dot{M}= \dot{M}/M_{BH}$ \citep{Merloni2010, Sun2015}, where $\rm \dot{M}$ is the accretion rate ($\rm \dot{M}=\frac{(1-\eta)L_{bol}}{\eta c^2}$, where $\eta=0.1$ is the assumed radiative efficiency of the accretion disk and c is the speed of light) and their specific galaxy stellar mass growth rate, sSFR. The distribution of the AGN used in our study in the  $\rm s\dot{M}/$sSFR$-$M$_{BH}/$M$_*$ space is presented in the top panel Fig. \ref{fig_specific}. Different colours and symbols correspond to different redshift bins. In agreement with \cite{Sun2015}, we find a strong anticorrelation between the two parameters, at all redshifts spanned by our dataset ($\rm r=-0.75\pm 0.12$). The horizontal dashed line indicates the $\rm s\rm \dot{M}/sSFR=1$, i.e., when the specific SMBH accretion rate and the sSFR are equal. The vertical dashed line denotes the average M$_{BH}/$M$_*$ ratio value found for our sample. 

The bottom panel of Fig. \ref{fig_specific} presents the distribution of systems that are in a SMBH growth dominant phase (red circles) and those that are in galaxy growth dominant phase (blue circles), in the $\rm s\dot{M}/$sSFR$-$M$_{BH}/$M$_*$ plane. We also define as "underweight", AGN that lie below the line that describes the best-fit on our data in the M$_{BH}$-M$_*$ plane (see black line in Fig. \ref{fig_mbh_vs_mstar}) and as "overweight", AGN that lie above this line (including, in both cases, its uncertainty of $\sim 0.4$\,dex). These sources are marked with a square (overweight) and a triangle (underweight) in the bottom panel of Fig. \ref{fig_specific}. We notice that the vast majority of galaxies with underweight SMBHs, have $\rm s\dot{M}/$sSFR$ >1$, that is their specific SMBH mass growth rate is higher than their specific stellar mass growth rate. In other words, their M$_{BH}$ is trying to catch up their M$_*$. On the opposite side, the vast majority of galaxies with overweight SMBHs have $\rm s\dot{M}/$sSFR$ <1$, that implies that the stellar mass growth rate is higher than the SMBH mass growth rate.


\section{Summary}

We used 687 X-ray luminous (median $\rm log\,[L_{X,2-10keV}(ergs^{-1})]=44.3$), broad line AGN, at $\rm 0.2<z<4.0$ (median $\rm z\approx 1.4$) that lie in the XMM-{\it{XXL}} North field. Their bolometric luminosities span nearly three orders of magnitude ($\rm 44<log\,L_{bol}\,(erg/s)<47$), while their BH and stellar masses range from $\rm 7.5<log\,[M_{BH}\,(M_\odot)]<9.5$ and $\rm 10<log\,[M_*(M_\odot)]<12$, respectively. Our goal was to study the co-evolution of the SMBHs and their host galaxies, over a wide redshift range. Our main results can be summarised as follows:

\begin{itemize}

\item[$\bullet$] The vast majority of the AGN in our dataset are in star-forming systems (Sect. \ref{sec_ms}). Their Eddington ratios range from 0.01 to 1, with a median value of 0.06 (Sect. \ref{sec_lbol_mbh}).

\item[$\bullet$] The M$_{BH}$ and M$_*$ are correlated. No statistical significant evolution of the M$_{BH}/$M$_*$ ratio is found with redshift, up to $\rm z=2$. The mean $\rm log($M$_{BH}/$M$_*)=-2.44$, with a 1$\sigma$ scatter of 0.61 (Sect. \ref{sec_mstar_mbh}).
    
\item[$\bullet$] Most of the AGN ($75\%$) are in a SMBH growth phase (AGN dominant phase). In systems that the galaxy mass growth is dominant, the M$_{BH}-$M$_*$ relation is significantly tighter compared to the galaxies that are in an AGN dominant phase. This could suggest that the growth of black hole mass occurs first, while the early stellar mass assembly may not be so efficient (Sect. \ref{sec_sfr_lbol}). 


\item[$\bullet$] We detect a strong anti-correlation between the  M$_{BH}/$M$_*$ ratio and the ratio of the specific SMBH and galaxy mass growth rates. Most of the AGN that their SMBH is classified as underweighted have $\rm s\dot{M}/$sSFR$ >1$, that is their specific SMBH mass growth rate is higher than their specific stellar mass growth rate. The majority of the AGN with overweighted SMBH have $\rm s\dot{M}/$sSFR$ <1$, which implies that their stellar masses are catching up their M$_{BH}$ (Sect. \ref{sec_specific}).

\end{itemize}

\begin{acknowledgements}
The author acknowledges support by the Agencia Estatal de Investigación, Unidad de Excelencia María de Maeztu, ref. MDM-2017-0765. This project has received funding from the European Union's Horizon 2020 research and innovation program under grant agreement n$^o$ 101004168, the XMM2ATHENA project.
This research has made use of TOPCAT version 4.8 \citep{Taylor2005}.

\end{acknowledgements}

\bibliography{mybib}
\bibliographystyle{aa}

\appendix

\end{document}